\documentstyle[aps,manuscript]{revtex}

\draft

\begin{document}
\title{Vortex distribution in neutron stars: gravitational effects.}
\author{H.Casini and R.Montemayor}
\address{Instituto Balseiro and Centro At\'{o}mico Bariloche\\
Universidad Nacional de Cuyo and CNEA\\
8400 S.C. de Bariloche, R\'{\i}o Negro, Argentina}
\maketitle

\begin{abstract}
Neutron stars are supposed to be mainly formed by a neutron superfluid. The
angular momentum is given by the vortex array within the fluid, and a good
account of the observable effects is determined by its coupling with the
crust. In this article we show that the gravitational field introduces
important modifications in the vortex distribution and shape. The inertial
frame dragging on the quantum fluid produces a decrease in the vortex
density, which for realistic models is in the order of 15\%. This effect is
relevant for neutron star rotation models and can provide a good framework
for checking the quantum effect of the frame dragging.
\end{abstract}


\section{Introduction}

In most quantum systems the gravitational effects are extremely small, but
there are some remarkable exceptions. One which has been already extensively
studied is the generation of inhomogeneities in the inflationary phase of
the universe, through the amplification and the transition to a classical
regime of the quantum fluctuation of the inflaton field\cite{kolb}. In this
case the relevance of the phenomena is related to the propagation of modes
with a wavelength comparable to the radius of the horizon of the universe,
i.e., the gravitational and the quantum characteristic lengths are of the
same order. A natural question which arises is if there are other cases
accessible from a phenomenological point of view, and the best candidate for
an affirmative answer is provided by the neutron stars. There is a consensus
not only about their existence but also about their direct relation with
pulsars\cite{shapiro}. As it is well known, the pulsar dynamics shows a very
interesting phenomenology closely related to neutron star models. Among the
most suggestive phenomena displayed by pulsars are the glitches, which are
closely linked to the internal dynamics of the star. The post-glitch
relaxation of the angular velocity has a large time scale, which is
interpreted as evidence of a crust containing a superfluid medium\cite
{alpar1}.The models usually assume a solid exterior crust, which is the
directly observable zone, and inner regions mostly constituted by superfluid
neutrons. The superfluid phase would be stable at the very high pressure in
the interior of the star\cite{pines1}. This image is supported by
theoretical computations\cite{alpar2} as well as observational evidence\cite
{pines2}. Given that a superfluid flow is irrotational, the angular momentum
of the star must be supported by vortex lines. They are coupled to the
exterior crust, and the different phenomena associated to the glitches are
related with this coupling\cite{epstein}. These models give a qualitatively
correct description for the glitches dynamics. In this article we show that
gravitational effects could be a very relevant ingredient to this
description. Although this article is concerned with neutron stars, our
conclusions can be applied to the rotation of boson stars.

In the following section we give a brief introduction to the
subject of the rotation of a superfluid star, using the weak field
approximation for the gravitational field. This approximation is useful for
understanding the phenomena, but it is not suitable for studying effects
that involve the gravitational field in the interior of the neutron stars
because of their high mass density. For this reason in Section III we
analyze the rotation of a superfluid star using a different and more
adequate approach. We consider there the exact expression for the
gravitational interaction, and expand the metric in powers of the angular
velocity. This is a small parameter for a neutron star, and thus we can
restrict the expansion to linear tems without a significant error. Finally,
in Section IV we study in detail the vortex distribution in the star
and the effects of the gravitational field on this distribution
using a covariant description for the superfluid.

\section{Weak field approximation to the superfluid star rotation}

To get an insight into the physical aspects of this problem we will discuss
it in the first place using a weak field approximation. Thus we will assume
a post-Newtonian metric\cite{weinberg1}:
\begin{equation}
g_{\mu \nu }=\eta _{\mu \nu }+h_{\mu \nu }\,\;\;,\;\;\,\,h_{\mu \nu }\ll 1\,,
\end{equation}
where the gravitational field is described by a vectorial potential, $\vec{h}
_i=h_{0i}$, and a scalar one, $\phi =\frac{h_{00}}2$. In this case the
Hamiltonian for a non relativistic particle becomes \cite{casini}:
\begin{equation}     \label{hamilt}
H=(1+3\,\phi )\frac{(\vec{p}+m\,\vec{h})^2}{2m}+m\,\phi \,.
\end{equation}
Except for the factor $(1+3\,\phi )$, which gives the red shift effect,
there is a clear similarity between this Hamiltonian an the corresponding
one for a particle in an electromagnetic field, where the electric charge $e$
should be replaced by the mass $m\,$, the vector electromagnetic potential
$\vec{A}$ by $-\vec{h}$, and the electric potential by the Newtonian one.
Extending the analogy one could suppose that a gravitational Meissner
effect, the expulsion of the field $\nabla \times \,\vec{h}$, is present
among the gravitational phenomena that take place in a superfluid star, as
was proposed in Ref.\cite{peng}. However, we can argue that this is not the
case on the basis of the Mach principle. It tells us that the field source
tries to impose its own rest frame to the other bodies through the
gravitational interaction. This implies that the gravitational effect on the
bodies has a paramagnetic nature independently of the substances they are
made of, and thus there is no such Meissner effect. This point will be
formalized later.

Let us now analyze the slow rotating star in this approximation. The
energy-momentum tensor of the fluid is $T^{\mu \nu }=(p+\rho )u^\mu u^\nu
+pg^{\mu \nu }$, where $p$ is the pressure and $\rho $ the energy density.
Expanding this tensor up to first order in the velocity and using the weak
field approximation to the Einstein equations in the harmonic gauge, the
equations for the field $\vec{h}$ become
\begin{eqnarray}
\Delta \,\vec{h} &=&16\pi G\,[(\rho +p)\,\vec{v}+\frac 12(3\rho +p)\,\vec{h}
]\,\,\,\,\,\,\,\,\,\,\,\,,\,\,\,\,\,\,\,\,\,\,  \label{ecu1} \\
\vec{\nabla}.\,\vec{h} &=&0\,.  \label{ecu2}
\end{eqnarray}

Based on the analogy with a superconductor, the Hamiltonian (\ref{hamilt})
implies that $\vec{\nabla}\times (\vec{v}+\vec{h})=0$. Furthermore, in a
superfluid star we have a stationary rotating fluid where $\vec{\nabla}.
\vec{v}=0$. These two conditions, together with Eq.(\ref{ecu2}) give
$\vec{v}=-\vec{h}$. Thus Eq.(\ref{ecu1}) leads to
\begin{equation}       \label{alfa}
\Delta \,\vec{h}=-8\pi G\,(\rho -p)\,\vec{h}\,.
\end{equation}
As $\rho >p,$ the paramagnetic character mentioned above is clearly stated.
A similar result was obtained in Ref.\cite{ciubotariu}, but there is a
difference because the authors do not take into account the term
proportional to $\vec{h}$ on the right hand side of Eq.(\ref{ecu1}). In the
present context this term has the same weight as the term proportional to
$\vec{v}$, since $\vec{v}=-\vec{h}$, and it should not be neglected.

The study of the rotation of the star reduces to the analysis of Eq.(\ref
{alfa}), which formally can be considered as a Schr\"{o}dinger equation for
$\vec{h}$ with a potential well proportional to $\rho -p$. In absence of
vortices the phases induced in the superfluid by the rotation and the field
$\vec{h}$ cancel exactly, and a rotational state of the star is given by a
bounded regular solution of this equation that nullifies $\vec{h}$ at
infinity. In general the existence of this solution implies a deep or long
potential well that leads to an unstable star that will collapse to a black
hole. At least this result shows that the weak field approximation is not
suitable for analyzing the possibility of a vortexless rotational state,
because non linear effects are important.

\section{Superfluid star rotation}

In this section we will analyze
the superfluid star rotation on the basis of an expansion of the metric
tensor $g_{\varphi t}$ in powers of the angular velocity $\Omega$,
which in the case of neutron stars can be considered as a small parameter.
This last point allows us to restrict the expansion up to linear terms in
$\Omega$\cite{hartle}. At first order on the angular velocity the metric
takes the form
\begin{equation}     \label{metrica}
ds^2=-e^{2\Phi }dt^2+e^{2\Lambda }dr^2+r^2(d\theta ^2+\sin {}^2\theta \
d\varphi ^2)-2r^2\sin {}^2\theta \ \omega \,d\varphi \,dt\,,
\end{equation}
where $\omega (r,\theta)$ represents the angular velocity that takes a free
falling object from infinity to the point $(r,\theta)$, and which
corresponds to the local inertial frames rotation with respect to the fixed
stars. The potentials $\Phi$ and $\Lambda$ are even functions of the
angular velocity, and therefore in this approximation they are the
non-perturbed functions of $r$ that can be computed directly from the
non-rotating stellar model. For stars at the end of the thermonuclear
evolution they can be determined from the equation of state and the central
pressure. The equations in this case are:
\begin{eqnarray}
&&m(r)=\int_0^r4\pi \rho dr\,,  \label{estruc} \\
&&\frac{dp}{dr}=-\frac{(\rho +p)\,(m+4\pi r^3p)}{r(r-2m)}\,, \\
&&\frac{d\phi }{dr}=\frac{m+4\pi r^3p}{r(r-2m)}\,, \\
&&e^{2\Lambda }=\frac r{r-2m}\,.
\end{eqnarray}
The remaining component of the metric, $\omega$, is given by the Einstein
equation corresponding to the component $R_{\varphi t}$ of the Ricci tensor.
To reduce this equation to a simple and suitable form, we can introduce the
angular velocity of the system, which can be precisely characterized by
$\Omega =u^\varphi /u^t$ in terms of the fluid velocity $u^\nu$. This
magnitude $\Omega$ represents the angular velocity measured by an observer
at rest with respect to the fluid. The minimal energy configuration, and in
consequence the stable one, has $\Omega$ constant, which restricts the
problem to the case of a uniform rotation. If we assume that the
constitutive matter behaves as a perfect fluid, $T_{\varphi t}$ is:
\begin{equation}
T_{\varphi t}=-r^2\sin {}^2\theta \,((\rho +p)\,\Omega \,-\rho \omega )\,.
\end{equation}
Besides, using the azimuthal symmetry of the rotating star, we can expand
$\omega $ as follows:
\begin{equation}
\omega (r,\theta )=\sum_{l=1}^\infty \omega ^l(r)\,(-\frac 1{\sin \theta }\,
\frac{dP_l}{d\theta })\,\,,
\end{equation}
where $P_l$ is the Legendre polynomial of degree $l$. With these ingredients
the perturbed Einstein equation becomes
\begin{eqnarray}
\omega ^l,_{rr}+(\frac 4r-\Lambda ^{\prime }-\Phi ^{\prime }\,)\,\omega
^l,_r+\frac 2r\,(\frac 1r+\Phi ^{\prime } &-&\Lambda ^{\prime }-\frac{l(l+1)
}{2r}e^{2\Lambda })\,\omega ^l  \nonumber \\
&=&16\pi e^{2\Lambda }(\frac 12(\rho +3p)\,\omega ^l\,-\,(\rho +p)\,\Omega
\,\,\delta _1^l)\,.  \label{algo}
\end{eqnarray}

The only information which remains to be introduced refers to the superfluid
state of the star matter. The superfluid is characterized by the curved
space-time covariant generalization of $\nabla \times \stackrel{\_}{\,\,v\,}
=0$, which is satisfied by a superfluid in absence of vortices in a flat
space-time. According to this the quadrivelocity must satisfy
\begin{equation}
\epsilon ^{ijkl}\xi _l^t(u_{j,k}-u_{k,j})=0\,,
\end{equation}
where $\xi $ is a time-like Killing vector. Due to the symmetries of the
system this relation reduces to $u_\varphi ,_r=0$, which we can rewrite as
\begin{equation}
\frac d{dr}(u^\varphi g_{\varphi \varphi }+u^tg_{t\varphi })=0\,.
\end{equation}

Substituting the metric components by the expressions which correspond to
(\ref{metrica}), we obtain that the quantity $r^2(u^\varphi -u^t\omega)$ is
independent of $r$ and therefore null. Hence we have
\begin{equation}       \label{igual}
\frac{u^\varphi }{u^t}=\Omega =\omega \,.
\end{equation}
We will reconsider this relation further on to include the presence of
vortices. By introducing it now in Eq.(\ref{algo}) and using (\ref{estruc})
we obtain the differential equation satisfied by $\omega$ in the superfluid
star:
\begin{equation}
\omega ^l,_{rr}+(\frac 4r-\Lambda ^{\prime }-\Phi ^{\prime }\,)\,\omega
^l,_r+\frac{2-l(l+1)}{r^2}e^{2\Lambda }\,\omega ^l=0\,.  \label{gamma}
\end{equation}

In this equation we can substitute $\omega =\frac{-h^\varphi }{r\,\sin
\theta }$, and develop up to the linear terms in the fields, thus recovering
Eq.(\ref{alfa}). For $\Lambda $ and $\Phi $ regular at the origin, i.e.,
nonsingular stars, the only solution with regular geometry is $\omega =0$.
Therefore there are no solutions for rotating superfluid stars without
vortices. A similar result was obtained in Ref.\cite{kobayashi} when
analyzing boson stars. In this case there is not an effective equation of
state and in that paper the boson star structure was computed using the
Klein-Gordon equation. Besides, the energy momentum tensor is not isotropic
because the radial and tangential pressures are not equal in the general
case. However, as in Eq.(\ref{algo}), only the radial pressure is relevant
and the resulting equation is again Eq.(\ref{gamma}). Due to the scalar
field coherence the equation $\Omega =\omega $ is satisfied, and hence the
energy momentum tensor is a function of $\omega $ only.

\section{Rotation in presence of vortices}

Because the superfluid star rotation cannot be achieved as a perturbation of
its fundamental state, we are going to study the rotation in presence of
vortices. In order to consider the vortex contribution to the superfluid
star dynamics in curved space, we will follow an approach analogous to the
one proposed by Weinberg\cite{weinberg2}. The neutron (and proton) field has
a U(1) global symmetry:
\begin{equation}
\Psi (x)\longrightarrow e^{i\Lambda }\,\Psi (x)\,,
\end{equation}
that leads to the baryonic number conservation. The superfluidity phenomena
is related to the formation of a condensate that spontaneously breaks this
global symmetry. This is similar to the superconductivity effect, where
there is a spontaneous breaking of the U(1) electromagnetic gauge symmetry
to Z$_2$. Based on the physical picture for the neutron star matter, where
neutron pairs have nonvanishing expectation values, we will assume that the
U(1) baryonic symmetry is broken to Z$_2$, the subgroup of transformations
with $\Lambda =0\,$ and $\Lambda =\pi $. The spontaneous breaking of the
symmetry leads to the existence of a Nambu-Goldstone excitation with zero
energy in the limit of vanishing momentum. The group transformation acting
on the Nambu-Goldstone boson is:
\begin{equation}
\phi (x)\longrightarrow \phi (x)+\Lambda \,.
\end{equation}

As $\phi $ parametrizes U(1)/Z$_2$ , $\phi $ and $\phi +\pi $ are taken to
be equivalent. The U(1) invariant density Lagrangian is a function of the
derivatives of $\phi $ and the U(1) fixed neutron fields $\tilde{\Psi}$, and
the baryonic Noether current is given by $j^\alpha =\frac{\delta L}{\delta
(\partial _\alpha \phi )}$ . The most general density Lagrangian allowed by
the symmetries is a nonlocal function of the field, but the nonlocality
extends over a range of the order of the penetration length of the
superfluid. Given that we are interested in the macroscopic fluid motion, we
will only consider the local terms in the density Lagrangian that
effectively describe the long range behavior. Such terms must be scalars and
should be constructed as a contraction of covariant quantities. The only
possible factors are the gradient of $\phi \,$, the metric (not the
curvature tensor because of the equivalence principle) and a number of fixed
tensors that characterize the field $\tilde{\Psi}$ and must satisfy the
requirements of spherical symmetry, since they are determined by the
unperturbed star. Let us call them $\lambda ,\lambda _\mu ^{(1)},\lambda
_{\mu \nu }^{(2)},..$. As was argued in Ref.\cite{weinberg2} the existence
of an equilibrium configuration with vanishing $\phi $ gradients rules out
the linear terms in its derivatives . The quadratic terms must not vanish
since the system has a spontaneous symmetry breaking. Therefore, the density
Lagrangian can be expanded as:
\begin{equation}
{\cal L}=\frac 12(f(\lambda )\,\phi _{,\nu }\phi ^{,\nu }+g(\lambda
)\,\lambda _{\mu \nu }^{(2)}\,\phi ^{,\nu }\phi ^{,\mu }+...)\,.
\end{equation}

We only need the quadratic terms because $\phi _{,\nu }$ is zero for the
static star, and therefore it is a first order quantity of the angular
velocity for a rotating one. Hence we have $j_\mu =f(\lambda )\,\phi _{,\mu
}+g(\lambda )\,\lambda _{\,\,\,\,\,\,\,\mu }^{(2)\,\nu }\,\phi _{,\nu }$.
Due to the spherical symmetry the tensor $\lambda _{\mu \nu }^{(2)}$ has the
same angular dependence as the metric. Thus $\lambda _{\,\,\,\,\,\,\,\varphi
}^{(2)\,\varphi }$ is a function of $r$ and we can write $j_\varphi
=n(r)\,\,\phi _{,\varphi }\,=n(r)\,\partial _\varphi \phi $. The current
component $j_\varphi $ can also be written for small velocities as
$n_0(r)\,u_\varphi$, where $n_0(r)=\frac 1{\sqrt{\gamma}}\frac{dn}{dV}$ is
the baryon number density in the fluid rest frame and $\gamma \,$ is the
determinant of the spacial metric tensor\cite{landau1}. From here we obtain
\begin{equation}
u_\varphi =\frac{n(r)}{n_0(r)}\partial _\varphi \phi =\frac 2{m^{*}(r)}
\partial _\varphi \phi \,,
\end{equation}
where $m^{*}=n_0/n$ is a scalar quantity of dimension one that depends only
on the fluid conditions. This parameter can be interpreted as the effective
mass of the quasiparticles from the equivalence principle, and for a neutron
star it takes a value of order of the neutron mass\cite{alpar1}\cite{wambach}
\cite{wiringa}. The Nambu-Goldstone field $\phi $ should be compared with the
 double the phase of the
Ginzburg-Landau wave function, which is the origin of the factor two in the
preceding equation\cite{weinberg2}.

On the other hand we have
\begin{equation}
\Omega =\frac{u^\varphi }{u^t}=\frac{u_\varphi g^{\varphi \varphi
}+u_tg^{\varphi t}}{u^t}\,.
\end{equation}
This expression becomes
\begin{equation}
\Omega =\frac{u_\varphi }{r^2\sin ^2\theta \,\,u^t}+\omega =\frac{2\;
\partial _\varphi \phi }{m^{*}\,r^2\sin ^2\theta }\,\,e^\Phi +\omega \,.
\end{equation}
when we keep up to the first order terms in $\omega $. The factor $e^\Phi $
gives the redshift of the effective mass. From the rotation symmetry, the
periodicity and the continuity of the $\phi $ field outside singularities we
obtain $\phi =n(r,\theta )\,\,\varphi /2$, where $n(r,\theta )\,$ is the sum
of the indexes of the field singularities held by a circular closed path in
a plane orthogonal to the rotation axis, centered on this axis, and
containing the point $(r,\theta )$. The $n(r,\theta )$ function is also
equal to the number of vortices surrounded by the closed path, because each
vortex has a topological number one to minimize the energy. In consequence
\begin{equation}          \label{doble}
\Omega =\frac{n(r,\theta )}{m^{*}\,r^2\sin {}^2\theta }\,\,\,\,e^\Phi
+\omega \,.
\end{equation}
This equation generalizes the expression (\ref{igual}) by including the
presence of vortices. It also generalizes the formula describing the angular
velocity and vortex density relation in flat space. From this point of view
the new ingredient is the second term on the left hand side of the equation,
which corresponds to the dragging of the inertial frames due to the
gravitational field. This relation shows that the phase introduced by the
vortices in the superfluid is equal to the sum of the kinetic phase plus a
gravitational phase. This phase is due to the metric that makes the
covariant and contravariant components different, unlike the electromagnetic
case and what can be interpreted from the weak field approximation where it
is introduced by a connection.

Given the angular velocity $\Omega $ and the star structure at rest, we can
state the vortex distribution in a superfluid star. First we compute $\omega
$ from Eq.(\ref{algo}), with the reasonable supposition that the vortex
interaction energy is negligible with respect to the rotation energy. Once
$\omega $ is obtained we can compute the vortex distribution and shape from
Eq.(\ref{doble}). The resulting vortex distribution minimizes the total
energy of the fluid, leading to a macroscopic motion equal to the
corresponding one without vortices. Since $\omega $ is a decreasing function
of the radius, the decrease in the vortex density with respect to a flat
space is greater at the axis of the star. Besides this, the gravitational
field produces a line vortex diffraction in such a way that these lines are
not parallel to the rotation axis, as in a flat space, except on the
equatorial plane. The angle $\beta $ between the rotation axis of the star
and the direction of the vortex lines at the point $(r,\theta )$ is given by
\begin{equation}
\sin \beta =\frac 12\frac{\kappa (r)\;\sin (2\theta )}{\left( 1+\kappa
(r)(\kappa (r)-2)\;\sin ^2(\theta )\right) ^{1/2}}\,,
\end{equation}
where $\kappa (r)=\frac r2\left( \frac 1{\Omega -\omega }\frac{d\omega }{dr}-
\frac 1{m*}\frac{dm*}{dr}+\frac{d\Phi }{dr}\right) $.

Outside the star we have $\omega (r)=\frac{2\,G\,J}{r^3}$ , where $J$ is the
total angular momentum. From here and applying Eq.(\ref{doble}) with $\theta
=\frac \pi 2$ and $r=R$, we get the total number of vortices
\begin{equation}
N=m^{*}(R)\,\,\left( \Omega R^2-\frac{2\,G\,J}{R^3}\right) \,\,e^{-\Phi
(R)}\,.
\end{equation}
Even if we do not know the star mass distribution, which is model dependent,
it is possible to give an upper bound for the average number of vortices
$\nu _g$ per invariant area unit, $dA=2\pi re^\Lambda dr$. Given that
$e^{\Lambda (r)}>e^{-\Phi (R)}$ for $r<R$. this upper bound is:
\begin{equation}
\nu _g=\frac NA=m^{*}\,\,\,(\Omega \,-\,\frac{2\,G\,J}{R^3})\frac{
R^2e^{-\Phi }}A\leq \frac{m^{*}}\pi \,\,\,(\Omega \,-\,\frac{2\,G\,J}{R^3}
)\,.
\end{equation}
Thus the relative decrease of the average vortex number density with respect
to the one at a flat space, $\nu _0=\frac{m^{*}\Omega }\pi $ , is:
\begin{equation}
\Delta =\frac{\nu _0-\nu _g}{\nu _0}\geq \frac{2\,G\,I}{c^2R^3}\,,
\end{equation}
where $I$ is the star moment of inertia. To clarify the meaning of the
formula we can put $I=M\,\bar{r}^2,$with $\,\bar{r}$ the radius of gyration,
and thus we get the decrease rate bound as $(\frac{R_s}R).(\frac{\bar{r}^2}
{R^2})$ , where $R_s=\frac{2\,G\,M}{c^2}$ is the Schwarzchild radius of the
star. For different neutron star models quoted in Ref.(\cite{wiringa}), this
lower bound varies from 11\% to 15\%. The vortex density decrease is greater
as the star is more relativistic. These results are interesting at least for
two reasons. On the one hand, they can be a significant ingredient for
analyzing the dynamics of neutron stars, provided the development of the
high pressure nuclear matter theory and pulsar models are accurate enough
for describing observations. On the other hand, they provide a framework for
the verification of the equivalence principle at the level of quantum
systems by checking the effect of frame dragging.

\section{Acknowledgments}

We are grateful to G. Zemba for very interesting and illuminating
discussions. This work was realized with a partial support from the Consejo
Nacional de Investigaciones Cient\'{\i}ficas y T\'ecnicas (CONICET),
Argentina.

\end{document}